\documentclass[12pt]{article}
\usepackage{graphicx}
\usepackage{color}
\usepackage{amsmath,slashed, amssymb}
\usepackage{fancyhdr}
\usepackage{hyperref}
\usepackage[titletoc]{appendix}
\numberwithin{equation}{section} 

\def\beq{\begin{eqnarray}}
\def\eeq{\end{eqnarray}}
\def\bea{\begin{eqnarray*}}
\def\eea{\end{eqnarray*}}

\def\centeron#1#2{{\setbox0=\hbox{#1}\setbox1=\hbox{#2}\ifdim
\wd1\rangle\wd0\kern.5\wd1\kern-.5\wd0\fi
\copy0\kern-.5\wd0\kern-.5\wd1\copy1\ifdim\wd0\rangle\wd1
\kern.5\wd0\kern-.5\wd1\fi}}
\def\ltap{\;\centeron{\raise.35ex\hbox{$\langle$}}{\lower.65ex\hbox{$\sim$}}\;}
\def\gtap{\;\centeron{\raise.35ex\hbox{$\rangle$}}{\lower.65ex\hbox{$\sim$}}\;}

\newcommand{\newc}{\newcommand}
\newc{\qbar}{{\overline q}}
\newc{\Kahler}{Kahler }
\newc{\deltaGS}{\delta_{\rm GS}}
\begin{document}
\begin{titlepage}
\begin{flushright}
{\large SCIPP 17/06\\
\large ACFI-T17-08 \\}
\end{flushright}

\vskip 2.2cm

\begin{center}

{\LARGE\bf Axions, Instantons, and the Lattice}

\vskip 1.6cm

{\large Michael Dine, Patrick Draper, Laurel Stephenson-Haskins, and Di Xu }
\\
\vskip 0.4cm
{\it $^{(a)}$Santa Cruz Institute for Particle Physics and
\\ Department of Physics, University of California at Santa Cruz \\
     Santa Cruz CA 95064  } \\
     ~\\
{\it $^{(b)}$Amherst Center for Fundamental Interactions, Department of Physics,\\ University of Massachusetts, Amherst, MA 01003
} \\
\vspace{0.3cm}

\end{center}

\vskip 4pt

%\vskip 1.5cm

\begin{abstract}

If the QCD axion is a significant component of dark matter, and if the universe was once hotter than a few hundred MeV, the axion relic abundance depends on the function $\chi(T)$, the temperature-dependent topological susceptibility.  Uncertainties in this quantity induce uncertainties in the axion mass as a function of the relic density, or vice versa. At high temperatures, theoretical uncertainties enter through the dilute instanton gas computation, while in the intermediate and strong coupling regime, only lattice QCD can determine $\chi(T)$ precisely.  
We reassess the uncertainty on the instanton contribution, arguing that it amounts to  less than $20\%$ in the effective action, or a factor of 20 in $\chi$ at $T=1.5$ GeV.
We then combine the instanton uncertainty
with a range of models for $\chi(T)$ at intermediate temperatures and determine the impact on the axion relic density. We find that for a given relic density and initial misalignment angle, the combined uncertainty amounts to a factor of 2-3 in the zero-temperature axion mass.

\end{abstract}

\end{titlepage}

\section{Introduction}

The axion remains a promising candidate for dark matter~\cite{abbottsikivie,preskillwilczekwise,dinefischler}, perhaps more so as the window for conventional WIMPs shrinks.
Searches for axion dark matter are underway~\cite{Rosenberg:2015kxa}, and there are proposals for future experiments which could conceivably
widen the search window substantially~\cite{grahamrosenberg}.

In the early universe, the axion begins to oscillate coherently when its thermal mass, $m_a(T)$, becomes comparable to the Hubble scale. The axion mass is related to the QCD topological susceptibility by 
\beq
m_a^2(T) f_a^2=\chi(T), ~~~~\chi=\int d^4x \langle F\tilde F(x)F\tilde F(0)\rangle_T=\partial_\theta^2 F(\theta,T),
\eeq
where, in the last expression, $F(\theta,T)$ is the $\theta$-dependent free energy.

At high temperatures, $F(\theta,T)$ can be calculated by standard instanton methods~\cite{gpy}. At low temperatures, $F(\theta,T)$ is known from chiral perturbation theory, and in fact converges rapidly to its $T=0$ limit below the confining phase transition~\cite{bonatia,borsanyia}. However, for plausible cosmologies and and a range
of axion parameters, it is the case that the axion starts to oscillate at intermediate temperatures, $T\sim$ GeV, where $\alpha_s$ is approaching strong coupling and neither calculation applies. Instead, one can hope to extract $\chi(T)$ from lattice QCD.

Recently there have been a number of papers reporting lattice calculations of $\chi(T)$ at temperatures above the critical temperature, both in pure gauge theory/quenched approximations~\cite{berkowitz, kitanorapidfalloff,borsanyia} and in QCD~\cite{kitano,sharma,borsanyinature,bonatia,bonatib,Trunin:2015yda,Burger:2017xkz}.   In some cases, discrepancies are found at high temperatures compared with the dilute instanton gas prediction. The free energies found in~\cite{sharma, borsanyinature} differ by about an order of magnitude from the leading-order semiclassical result above the GeV scale, while the computation and extrapolation obtained in~\cite{bonatia,bonatib} (see also~\cite{villadoro}) differs by many orders of magnitude (although systematic effects are not under control in the extrapolation, and thus the level of compatibility with the controlled high-temperature computations in~\cite{sharma, borsanyinature} is unclear.) %To compare 
These results suggest a level of uncertainty in the microscopic parameters (the zero temperature axion mass $m_a$ and possibly the misalignment angle $\theta_0$) required to achieve a given relic density $\Omega$.

Here we will assess the theoretical uncertainty on $\Omega(m_a,\theta_0)$ in analytic computations and compare with lattice results. Our analysis has two prongs. First, in Sec.~\ref{errorestimate}, we reexamine the uncertainty on the leading-order semiclassical result for the free energy above the GeV scale. We consider a number of possible sources of error, the most important of which are likely to be corrections to the effective action that shift the IR cutoff on instanton sizes.  However, in contrast to suggestions in the literature, we argue that infrared divergences that plague ordinary finite-temperature perturbation theory are not numerically relevant in the instanton effective action, and that the size of higher-order corrections can be reasonably estimated. 
As a result, the topological susceptibility
has known low and high temperature asymptotics which appear compatible with~\cite{sharma, borsanyinature}. 
Then, in Sec.~\ref{chimodel}, we introduce a family of models for the topological susceptibility that interpolate through the region where neither analysis is reliable. In Sec.~\ref{axioncosmology}, we compute the relic density over this range of models, including the instanton uncertainty at the high temperature boundary and uncertainties in QCD parameters. In this way we determine the sensitivity of $\Omega(m_a,\theta_0)$ to theoretical uncertainties. We find the sensitivity is  limited, and overall 
the axion mass prediction from analytical methods appears robust at the level of a factor of $2-3$. 
We comment on the implications of these results in Sec.~\ref{conclusions} and conclude.

\section{Theoretical Uncertainties on the Instanton Contribution to the Free Energy}
\label{errorestimate}

\subsection{The Standard Computation}

At high temperatures, the $\theta$ dependence of the free energy is controlled by instantons~(\cite{gpy}, henceforth GPY). Classically, even at finite temperature, there are instantons of all scale sizes.  But at one loop, there are two
sources of scale invariance violation:  the usual ultraviolet divergences familiar in the zero temperature
theory, and the finite temperature itself.  Both correct
the effective action, rendering finite the scale size integral both at small and large $\rho$.

Heuristically, the latter effect is associated with the effective mass of the $A_4$ field,
\beq
m_D^2 = {1 \over 3} (g^2 T^2)(N + {N_f \over 2}).
\eeq
GPY note that
a term in the effective action ${1 \over 2 g^2} m_D^2 A_4^2$ gives rise to a correction to the instanton action, for $\rho \gg T^{-1}$ (and $\alpha_s(\rho^{-1})\ll1$), proportional to $\rho^2$:
\beq
\int d^4 x { 1 \over 2 g^2} m_D^2 A_4^2 ={\pi^2 \over 2 g^2} m_D^2 \rho^2.
\label{appxaction}
\eeq
Note the $g^{-2}$ in front of $m_D^2$, reflecting the $1/g^2$ in front of the whole action, and the fact that the
actual screening length is of order $1 \over gT$.   If this were the complete
result for the correction to the effective action, the $\rho$ integration for the free energy would take the form, in the case of three
flavors,
\beq
F(T) \propto m_u m_D m_s \int {d\rho \over \rho^2} (\Lambda \rho})^9 e^{-{3} \pi^2 \rho^2 T^2.
\eeq
The integral is finite, and dominated by $\rho \sim (\pi T)^{-1}$.  

Since the dominant scale is of order  $ T^{-1}$,  the effective action cannot be expanded in powers of $\rho$;
in a derivative expansion of the background field effective action, 
terms of the form
\beq
{g^2 \over T^{n-2}} A_4(\vec x) \partial_{i_1}\dots \partial_{i_n } A_4(\vec x)
\eeq 
are all of the same order, $g^2 T^{2}$, in the instanton background.

GPY indeed computed
the full one-loop determinant~\cite{gpy}.  At small $\rho$, in particular, the above expression for the action is modified:
\beq
\delta S =  {1 \over 3} \pi^2 \rho^2 T^2 (2N + N_f) - {1 \over 18} \pi^2 \rho^2 T^2 (N-N_f).
\eeq
For $N_f =0$, for example, this is not parametrically smaller than the Debye screening term, though it is numerically smaller.
At one loop, the complete expression for the free energy in the presence of a single instanton is given by~\cite{gpy} 
\beq
F(\theta,T) = -\int {d\rho \over \rho^5}\left ( {4 \pi^2 \over g^2} \right )^{2N}  e^{-{8 \pi^2 \over g^2(\rho)}} C_N \prod_{i=1}^{N_f}\left( \xi \rho m_i\right) e^{-1/3 \lambda^2 (2N + N_F)
-12 A(\lambda)[1 + {1 \over 6} (N-N_f)]+i\theta} 
\label{eq:fullF}
 \eeq
 where 
 \begin{align}
 A(\lambda)&= -{1 \over 12} \ln (1 + \lambda^2/3) + \alpha(1 + \gamma \lambda^{-2/3})^{-8} \\
 \lambda = \pi \rho T~~~C_N&= 0.097163;~~~~\xi = 1.3388~~~\alpha = .01290~~~\gamma = 0.1586\;
 \end{align}
 and $N_F=3$ in temperature regimes where three quarks are excited. 
At a temperature of $T=1.5$ GeV and using a renormalization scale $\mu=T$,  we obtain 
\begin{align}
F_0(1.5)=-3.7\times10^{-14} {\rm~GeV}^{-4}  
\label{eq:F0}
\end{align}
where the subscript indicates $\theta=0$. Here we have used the program RunDec~\cite{rundec1,rundec2,rundec3} to obtain 
$\alpha_s(1.5~\rm{GeV})\simeq 0.345$ with three active flavors.

 This computation of the free energy is subject to certain theoretical uncertainties, including higher-order corrections sensitive to the UV cutoff, parametric uncertainties on $\alpha_s$, effects of heavier quarks, and higher-order corrections that modify the infrared cutoff on $\rho$. In the next subsection, we estimate the uncertainties from the first three of these sources. In our view there has been some confusion in the literature about the uncertainty associated with corrections to the $\rho$ cutoff, which is plausibly the dominant source of uncertainty. We therefore devote a separate subsection to this source.

\subsection{UV-Sensitive Corrections, Heavy Quarks, and Parametric Uncertainties}
In the previous section, we evaluated the one-loop expression for $F$ at $T=1.5$ GeV with $\mu=T$ and three flavors. Let us comment on a few of the knobs we can turn in this calculation to obtain estimates of theoretical uncertainty.

\begin{itemize}
\item Because the dominant instanton size is of order $(\pi T)^{-1}$,  $\mu=\pi T$ is another natural choice for the renormalization scale. 
\item At $T=1.5$ GeV, $\pi T$ is substantially above the charm threshold and near the bottom quark mass. We might therefore include at least the charm quark in the free energy.
\item A complete two-loop calculation of $F$ is not available at present. In some places in the literature, UV-divergent two-loop corrections to the free energy are incorporated using renormalization group considerations, as discussed in~\cite{morris}. These corrections are generally written as powers of $\alpha_s(\rho^{-1})/\alpha_s(\mu)$ in the $\rho$ integrand. However, for $\mu$ of order $T$ (or $\pi T$), there is no justification for including some two-loop corrections and not others. Without performing an actual two-loop computation of the $\theta$-dependent part of the free energy, the only principled approach is to use the complete one-loop expression with $\mu$ of order $T$ (or $\pi T$). However, the ``UV two-loop" computation might instead be useful as an uncertainty estimator.
\end{itemize}

We therefore recompute the free energy using $\mu=\{1,\sqrt{\pi},\pi\}\times T$, three or four active flavors, and including or not including the UV-divergent two-loop corrections. For the change in renormalization scale, we again use RunDec to accurately determine $\alpha_s(\mu)$ with different numbers of active flavors. The two-loop corrections are incorporated by running $\alpha_s$ in the exponent to $\mu=\rho^{-1}$ at two-loop order and running the quark masses and the coupling in the prefactor to $\mu=\rho^{-1}$ at one-loop order. For example,
\begin{align}
e^{-\frac{2\pi}{\alpha_s(\mu)}}\rightarrow e^{-\frac{2\pi}{\alpha_s(\mu)}}(\mu\rho)^{b_0}\left(\frac{\alpha_s(\rho^{-1})}{\alpha_s(\mu)}\right)^{2b_1/b_0}\;
\end{align}
where $b_0=9$ and $2b_1/b_0=32/9$ for $N_F=3$, and $\alpha_s(\rho^{-1})$ is determined from $\alpha_s(\mu)$ at one-loop order.

Results are reported in Table~\ref{tbl:freeenergy}. The largest value for the free energy is obtained in the three-flavor scheme with $\mu=T$, adding the partial two-loop terms. This is not a surprise, since the included two-loop terms correspond entirely to  running from $T$ to $\rho^{-1}\sim\pi T$. However, it is likely an overestimate of the correction; if we use $\mu=\pi T$ in the same computation, the partial two-loop result is smaller and much closer to the complete one-loop result. In reality, the complete two-loop result is likely to involve a mixture of scales, motivating the choice $\mu=\sqrt{\pi}T$. 
We observe that (excluding the $3F, 2L, \mu=T$ result), the envelope of the values is contained within the $\mu=\sqrt{\pi}T$ calculations, corresponding to  an ${\cal O}(1)$ uncertainty the free energy,
\begin{align}
\frac{\Delta F_0(1.5)}{F_0(1.5)}\simeq 1\;.
\label{eq:uncUV}
\end{align}

\begin{table}[t!]
\centering
\begin{tabular}{|l|l|}
\hline
3F, 1L, T             & 3.6 \\ \hline
3F, 2L, T             & 10 \\ \hline
3F, 1L, $\sqrt{\pi}$T & 4.9 \\ \hline
3F, 2L, $\sqrt{\pi}$T & 7.2 \\ \hline
4F, 1L, $\sqrt{\pi}$T & 3.2 \\ \hline
4F, 2L, $\sqrt{\pi}$T & 5.2 \\ \hline
3F, 1L, $\pi$T        & 6.0 \\ \hline
3F, 2L, $\pi$T        & 5.5 \\ \hline
4F, 1L, $\pi$T        & 4.0 \\ \hline
4F, 2L, $\pi$T        & 3.8 \\ \hline
\end{tabular}
\caption{The instanton-induced free energy  in units of $-10^{-14}$ GeV$^{-4}$ at $\theta=0$ and $T=1.5$. Rows correspond to a variety of computations: (3F,4F) = three or four light flavors; (1L,2L) = one-loop complete or partial two-loop; (T,$\sqrt{\pi}$T,$\pi$T) = renormalization scale. }
\label{tbl:freeenergy}
\end{table}

We can also estimate a ``parametric" uncertainty stemming from experimental uncertainty in $\alpha_s$. Using the 1-sigma error bar on $\alpha_s(m_Z)$, running down to $\mu=\pi\times 1.5$ GeV and converting to the three-flavor scheme with RunDec~\cite{rundec1,rundec2,rundec3}, we obtain less than $2\%$ uncertainty in $\alpha_s$.\footnote{Uncertainty from higher order corrections to the running of $\alpha_s$ are extremely subdominant.} This results in an uncertainty in $F_0(1.5)$ of about a factor of 2, similar to the uncertainty from UV-sensitive corrections.

The lattice result for the topological susceptibility obtained in Ref.~\cite{borsanyinature} corresponds to $F_0(1.5)\approx -4\times 10^{-13}  {\rm~GeV}^{-4} $, which lies outside the uncertainty range that we estimate from these sources. Similar conclusions were drawn in the lattice studies~\cite{sharma,Burger:2017xkz}. In~\cite{villadoro,sharma}, it was suggested that the uncertainty in the 1-loop instanton computation arising from higher order terms could actually be much larger, associated with infrared divergences in QCD perturbation theory at finite temperatures and with large shifts in the Debye screening length. We now turn to corrections of this type.

\subsection{Corrections to the IR Cutoff on $\rho$ and Infrared Sensitivity}
\label{dilutegas}
As discussed previously, the IR cutoff on the instanton size can be \emph{qualitatively} associated with the Debye mass term in the effective action. In the perturbative vacuum, the Debye mass does receive large corrections beyond leading order~\cite{rebhan,arnoldyaffe}.  
A simple, heuristic understanding of these corrections can be obtained by considering $\Pi_{44}$ as a function of
(spatial) momentum, $\vec q$, for small $\vec q$.  There are a variety of effects, but already at one loop, for example, there
is a contribution to $\partial \Pi/\partial q^2$ that 
diverges linearly as $m_D \rightarrow 0$ at $q=0$.  The linear IR divergence is cut off by the leading-order $m_D$, leaving a weaker logarithmic IR divergence cut off by the nonperturbative magnetic mass. The NLO Debye mass has the form~\cite{rebhan,arnoldyaffe}:
\beq
m_D^2 = (m_D)_0^2 + {2 N g^2 \over 4 \pi} T (m_D)_0 \ln(m_D/g^2T)+\dots.
\label{eq:deltamd}
\eeq
The NLO correction is of order $g^3$, signaling a breakdown of the perturbation expansion.   
It has been suggested~\cite{villadoro,sharma} that the uncertainty on $\chi(\theta, T)$ might be much larger than estimated in the previous section, due to the presence of such IR divergences.\footnote{It should be noted that the existence of such corrections is not connected with the presence 
of light fermions.  In particular, fermions do not introduce infrared divergences at high temperature.
Therefore, even lattice studies focusing on the size of corrections in the pure gauge theory are of interest.}

In the instanton computation, there is both a question of principle and a question of numerics. We have seen that it is not low spatial momenta that are relevant in the instanton background, but momenta of order $k\sim 1/\rho\sim T$. 
%, and the effective action should not be thought of in terms of a derivative expansion. 
%The contribution to the instanton effective action $S_{eff}$ associated with $\Pi_{44}$ is controlled by these momenta,  
Consequently, for the dominant semiclassical configurations with $\rho\sim 1/T\ll \Lambda$, the IR divergences in $\Pi_{44}$ are cut off at $T$ in the instanton effective action $S_{eff}$. 
%comes from $k\sim 1/\rho\sim T$, 
%Put another way, at weak coupling, the IR divergences in $\Pi_{44}$ are cut off at $T$ in the dominant semiclassical configurations, 
%and individual diagrams contributing to $\Pi_{44}$ 
The corrections to $S_{eff}$ from individual diagrams are then well-behaved and proportional to $g(T)^2 T^2$. Thus, as a matter of principle, IR divergent corrections to $m_D$ in the perturbative vacuum do not indicate a loss of perturbative control or a significant source of uncertainty in the instanton computation of $F(\theta,T)$. 

However, until $T$ is extremely large, $g(T)$ is ${\cal O}(1)$ in QCD, and there is no parametric separation between $m_D$ and $T$. Therefore Eq.~(\ref{eq:deltamd}), valid in the perturbative vacuum, might still be used as an estimate for the typical size of corrections to the effective action in the instanton background. Numerically, it gives rise to
\begin{align}
\frac{(m_D)_1}{(m_D)_0}\simeq 0.6
\label{eq:md1md0}
\end{align}
at $T=1.5$ GeV. The instanton-induced free energy scales approximately as the $7^{\rm th}-8^{\rm th}$ power of the infrared cutoff on $\rho$, so from Eq.~(\ref{eq:md1md0}) we are led to associate an uncertainty in the free energy due to two-loop finite temperature corrections,
\begin{align}
\frac{\Delta F_0(1.5)}{F_0(1.5)}\simeq 20\;.
\label{eq:uncdeb}
\end{align}

We emphasize that a correction to the free energy of this size does {\emph{not}} reflect a breakdown of the semiclassical analysis at this order. Organizing the instanton effective action as 
\begin{align}
S_{inst}=S_0+S_1+S_2+\dots,
\end{align}
at leading order, the action is
\begin{align}
S_0=\frac{8\pi^2}{g^2}\simeq 17
\label{eq:action}
\end{align}
at $T = 1.5$ GeV. A shift in the free energy of order Eq.~(\ref{eq:uncdeb}) corresponds to 
\begin{align}
\frac{S^{\rm \tiny Debye}_2}{S_0}\lesssim 0.2,
\end{align}
a controlled correction to the effective action. More generally, 
%the Debye mass is only one of many terms in the effective action that contribute at the same order to the IR cutoff on $\rho$, and 
we could estimate terms in the series by the three-dimensional loop factor, which is of order
\beq
\lambda = {N g^2 (T) \over (4 \pi)^{3/2}}.
\eeq
At $T = 1.5$ GeV,  $\lambda = 0.3$. Two-loop corrections to the effective action would then be expected to be of order 
\begin{align}
\frac{S_2}{S_0}\simeq \lambda^2 = 0.1\,
\end{align}
consistent with the Debye estimate. In other words, there is no reason to expect 
%a formal breakdown of perturbation theory, or 
{\emph{arbitrarily}} large corrections. The action is exponentiated in the free energy, leading to the order-of-magnitude uncertainty estimate in Eq.~(\ref{eq:uncdeb}).

There is also the question of actual infrared divergent contributions to
the instanton action. These are associated with  low momentum $\vec A$ fields and corrections to the effective action involving no background fields. In the zero-instanton
sector, such infrared divergences arise in the free energy first at four-loop order. It is
believed that they are cut off at a scale of order $g^2 T$, the presumed mass gap of the three
dimensional gauge theory. The typical diagram involves six vertices connected
by propagators, and the divergence arises when all vertices are well-separated. 
A computation at high order in the instanton background is complex, but for $\rho$ of order $T^{-1}$, the infrared divergence should be similar. At zero
temperature, the propagators are known~\cite{creamer,Levine:1978ge}, and at distances large compared to $\rho$, they
are close to free-field propagators. At finite temperatures, when all coordinates except
$x_4$ are large compared to $T^{-1}$ and $\rho$, we expect something similar, leading to an
infrared divergent correction at the same order as at zero temperature. At 1.5 GeV,
this suggests a perturbatively incalculable correction to the instanton action at the $1\%$ level.

%As emphasized previously, the Debye mass is only one of many terms in the effective action that contribute at the same order to the IR cutoff on $\rho$. More generally, we could estimate the size of typical finite-temperature corrections to obtain an uncertainty. The three-dimensional loop factor is of order
%\beq
%\lambda = {N g^2 (T) \over (4 \pi)^{3/2}}.
%\eeq
%At $T = 1.5$ GeV,  $\lambda = 0.3$. 
%Two-loop corrections to the effective action would then be of order 
%\begin{align}
%S_2/S_0\simeq \lambda^2 = 0.1\,
%\end{align}
%consistent if slightly smaller than the Debye estimate.

%Exponentiating, we obtain another uncertainty estimate,
%\begin{align}
%\frac{\Delta F_0(1.5)}{F_0(1.5)}\simeq 5\;,
%\label{eq:unctwol}
%\end{align}
%compatible with if somewhat smaller than the Debye estimate in Eq.~(\ref{eq:uncdeb}).

In summary,  IR divergences do not appear relevant to the instanton computation, and semiclassical analysis is under sufficient theoretical control to admit uncertainty estimates.
% large corrections are still possible because $g$ is ${\cal O}(1)$ and the  high temperature gauge theory is effectively three-dimensional.  
 Absent a complete 2-loop computation, we will take the perturbative Debye mass correction, Eq.~(\ref{eq:uncdeb}), as a \emph{conservative} estimate of the uncertainty in the $\theta$-dependent free energy. UV cutoff-sensitive corrections and uncertainties in $\alpha_s$ are expected to be subdominant to the finite-$T$ corrections to the effective action, and Eq.~(\ref{eq:action}) indicates that dilute gas corrections are expected to be negligible.

%Moreover, we emphasize that corrections cannot be \emph{arbitrarily} large: while an order of magnitude in the free energy is possible, it arises because the action is exponentiated, and corrections to the action are under control. 

We therefore know with some confidence the range of possible behaviors for the axion potential both at temperatures below the critical temperature ($\sim 150$ MeV for $N_f=3$) and at temperatures a few GeV and above. 
These boundary properties constrain the behavior in the intermediate range of temperatures, which happen to lie where the axion begins to oscillate in conventional scenarios.
For this reason lattice computations (that successfully reproduce the high temperature behavior) can be of value.
On the other hand, as we will describe below,
if we simply assume a smooth interpolation between the two regimes, the axion relic
density is not very sensitive either to the form of the interpolation or the uncertainty in the high energy semiclassical computation. 

In closing this section, we note that there have been arguments that the behavior of $\chi$ is drastically different at high temperatures than
the semiclassical result, even turning off exponentially rapidly with temperature~\cite{tcohen,aoki,Azcoiti:2016zbi,Azcoiti:2017jsh}.  We will not address this possibility further here, but it is certainly true that in such a circumstance substantially different axion relic densities can be obtained~\cite{kitanorapidfalloff,Azcoiti:2016zbi,Azcoiti:2017jsh}.

\section{$\chi(T)$ At Intermediate Temperatures}
\label{chimodel}
We have argued that we know the high-temperature behavior of $\chi$ to about an order of magnitude. 
At scales below $1~{\rm GeV}$, the coupling rapidly becomes strong, and other methods are needed to determine the axion mass.

At very low temperatures, the $\theta$ dependence of the vacuum energy is known reliably from current algebra,
\beq
F(\theta,0) = -3.6 \times 10^{-5} ~{\rm GeV}^4\cos(\theta).
\eeq
Finite temperature lattice computations indicate that the topological susceptibility ,
\begin{align}
\chi(T)=\frac{\partial^2 V(T)}{\partial\theta^2}\bigg|_{\theta=0},
\end{align}
 is near its zero-temperature ChPT value at temperatures of order 100 MeV, and remains approximately equal to it until at least the chiral phase transition near 150 MeV~\cite{borsanyia}. Beyond this scale, only lattice computations can accurately determine $\chi(T)$, and at present there are varied results in the literature. 

However, for the purposes of computing the axion relic density, it turns out to be sufficient to consider simple models that interpolate between the ChPT and instanton regimes.  We will adopt the following class of models for $F(\theta,T)$:
\beq
F(\theta,T) =
\Bigg\{ \begin{matrix}
- \chi(0)\cos\theta, &0<T<T_2 \\
  -\chi(T_0) \left ({T_0 \over T} \right )^n \cos \theta, &T_2<T<T_0  \\
   -\chi(T_0)\left ({T_0 \over T} \right )^8 \cos \theta, &T>T_0
\end{matrix}  
\label{freeenergyinterpolation}
\eeq
Here $T_0$ is the ``anchor point" for the instanton regime. The results of~\cite{borsanyia} suggest that the slope of $\chi$ is instanton-like down to temperatures a few times $T_c$; however, to maintain a minimal uncertainty in the semiclassical computation, we fix $T_0=1.5$ GeV. As discussed below, our modeling still includes the possibility of instanton-like slopes at lower $T$. We will vary $\chi(T_0)$ within the uncertainty on the instanton computation. $T_2$, the anchor point for the ChPT regime, is related to $T_0$ and the slope of the power law in the model by
\beq
T_2^n= T_0^n\times\frac{\chi (T_0)}{\chi(0)}.
\eeq
We vary $n$ such that $T_2$ varies between $100$ and $500$ MeV. Given $T_0=1.5$ GeV, values of $T_2$ of order $150$ in fact correspond to $n\simeq 8$, equivalent to assuming instanton-like behavior persists significantly below $T_0$.  Larger values for $T_2$ above the critical temperature $T_c$ are not based on physical considerations, but instead are included to partially accommodate the lattice results of~\cite{bonatia,bonatib}, which found very \emph{shallow} falloff of $\chi(T)$ above the chiral phase transition. This behavior is then approximated in the models of Eq.~(\ref{freeenergyinterpolation}) for larger $T_2$ by zero falloff until $T_2$. However, Ref.~\cite{bonatia,bonatib,villadoro} extrapolated the shallow power law behavior up to high temperatures, leading to values for $F_0$ many orders of magnitude different from the semiclassical result.  Our insistence on reaching instanton behavior by 1.5 GeV (within the uncertainty~(\ref{eq:uncdeb})) requires an even \emph{steeper} power law to set in above $T_2$ when $T_2 \gg T_c$. Numerically, $n$ will fall in the range $7-20$, with the lower values corresponding to lower values of $T_2$.

 \section{Axion Relic Density from Misalignment}
\label{axioncosmology}
We can now assess the sensitivity of the axion relic density to uncertainties in $\chi(T)$, including both the uncertainties in the instanton computation and the range of models for the behavior at intermediate temperatures. In the figures below we will numerically integrate the equation of motion,
\beq
\ddot a + 3 H \dot a + V^\prime(a) = 0.
\eeq
However, for qualitative purposes, a good approximation is obtained by treating the axion as frozen until a temperature $T_{osc}~$\cite{turnerreview,kolbturner}:
\beq
m_a(T_{osc}) = 3 H(T_{osc}).
\label{eq:maH}
\eeq
At this point, the axion begins to oscillate with a time (temperature) dependent mass.  Approximating the energy density by
\beq
\rho(t) = {1 \over 2} \dot a^2 + {1 \over 2} m_a^2(T) a^2,
\eeq
one can show that it evolves with temperature as
\beq
\rho(T) = \rho(T_{osc})\left ( {R^3(T_{osc}) \over R^3(T) }\right ){m_a(T) \over m_a(T_{osc})}.
\eeq
Within the range of $m_a(T)$ that we consider, $T_{osc}$ is always less than the instanton anchor point $T_0$. Therefore, Eq.~(\ref{eq:maH}) can be solved by substituting the intermediate power-law behaviors for $m_a(T)$; the instanton asymptotics constrain the range of intermediate power laws considered.

\begin{table}[t!]
\begin{center}
\begin{tabular}{|c|c|c|c|}
\multicolumn{4}{c}{$\Omega$} \\
\hline
$n$ & $\chi_0=1/10$ & $\chi_0 = 1$ & $\chi_0 = 10 $ \\
\hline
8 & 0.22 & 0.18 & 0.15\\
14 &  0.18 & 0.16  & 0.14\\
 20 & 0.17 & 0.15  & 0.14\\
 \hline
\end{tabular}
\caption{Axion relic density as a function of model parameters as computed with the approximate formula~(\ref{omegaresult}). Here $m_a(0)=30~\mu$eV, $\chi_0$ is given in units of $3.7\cdot 10^{-14}~{\rm GeV}^4$, and the misalignment angle is set to the value appropriate for post-inflationary breaking of the Peccei-Quinn symmetry, $\theta_0=2.16$.}
\label{omegatable}
\end{center}
\end{table}

The relic density $\Omega_a $ can then be expressed as a function of the parameters $\chi(T_0)$ and $n$ (or $T_2$).  The result is:
\begin{align}
 \Omega_{axion} = 0.13\times(7.3)^{\frac{2}{4+n}}\left(\frac{m_a}{30~\mu{\rm eV}}\right)^{-\frac{6+n}{4+n}}\left(\frac{\chi_0(1.5)}{3.7\cdot 10^{-14}~{\rm GeV}^4}\right)^{-\frac{1}{4+n}}\left(\frac{\theta_0}{2.155}\right)^2
\label{omegaresult}
\end{align}
where 
$\theta_0$ is the initial misalignment angle and $m_a$ is the zero-temperature axion mass. 
Taking, for example, $m_a=30~\mu$eV and a few values for $\chi_0(1.5)$ and $n$ gives the results in Table~\ref{omegatable}.
Alternatively, for fixed $\Omega_{axion} = \Omega_{\rm DM}$, we obtain
\begin{align}
\left(\frac{m_a}{30~\mu{\rm eV}}\right) = 0.51\times\left(2.4\right)^{\frac{6}{6+n}}\left(\frac{\chi_0(1.5)}{3.7\cdot 10^{-14}~{\rm GeV}^4}\right)^{-\frac{1}{6+n}}\left(\frac{\theta_0}{2.155}\right)^{\frac{8+2n}{6+n}}\left(\frac{\Omega_{DM}}{0.25}\right)^{-\frac{4+n}{6+n}}\;.
\end{align}

\begin{figure}[t!]
\begin{center}
\includegraphics[width=0.55\linewidth]{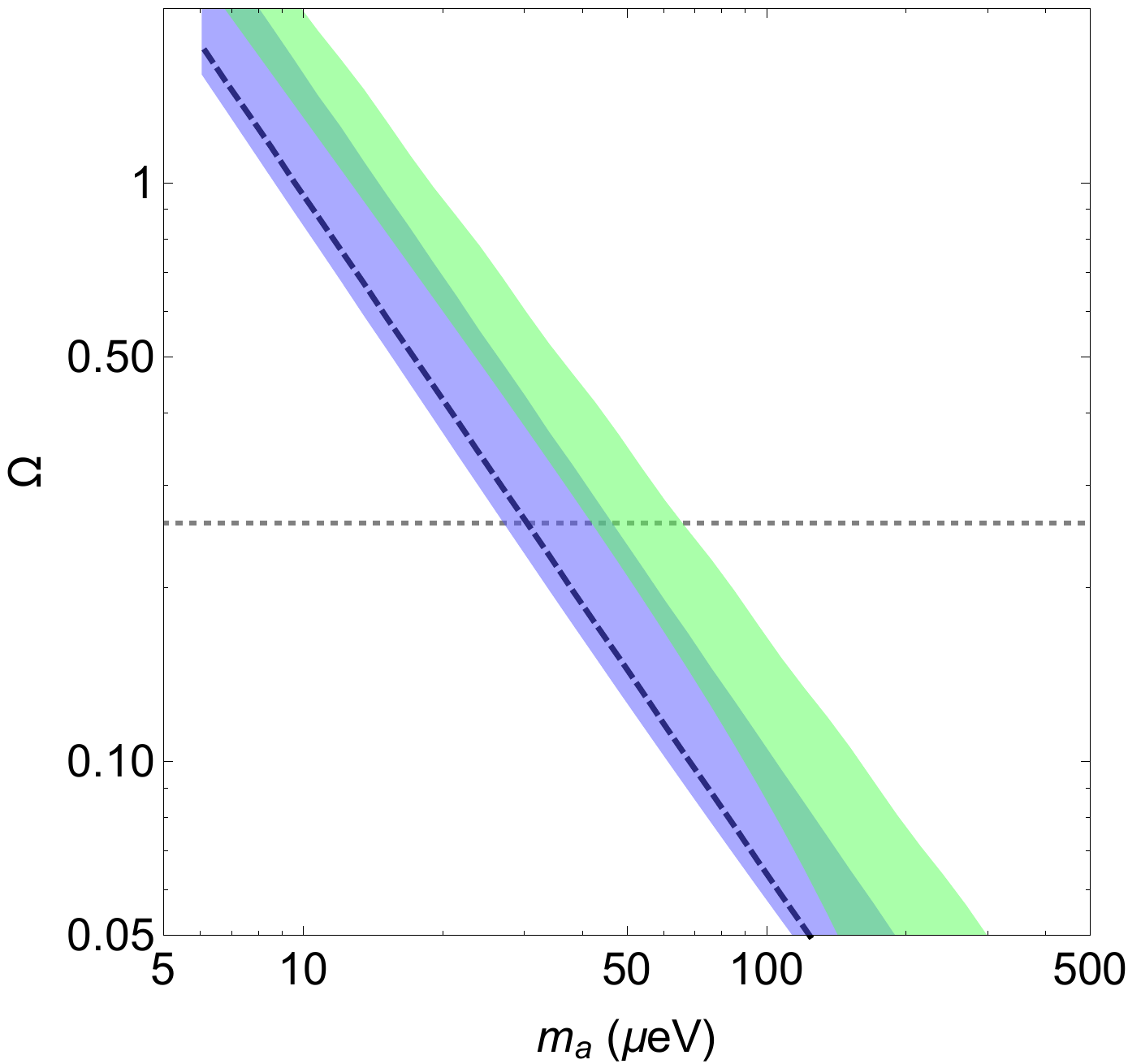}
\caption{Axion relic density from misalignment in the post-inflationary scenario. Colors correspond to different models for the temperature-dependent free energy between the dilute gas at high temperatures and chiral perturbation theory at low temperatures. Specifically, the blue (green) band sets the anchor point for ChPT at $T_2=100~(500)$ MeV. The width of each band reflects the uncertainty in the instanton computation of the free energy used as an anchor at $T=1.5$ GeV, $F_0\rightarrow (1/20,20)\times F_0$, c.f. Eqs.~(\ref{eq:F0}),(\ref{eq:uncdeb}). The dashed line corresponds to the value of $F_0(1.5)$ obtained in the lattice calculation of Ref.~\cite{borsanyinature}.} 
\label{fig:relic1}
\end{center}
\end{figure} 

Eq.~(\ref{omegaresult}) indicates that the relic density is substantially insensitive to the magnitude of the free energy at high temperatures; for $T_2=150$ MeV ($n\sim 8$), $\chi_0$ enters to the $-\frac{1}{12}$ power. Therefore, sizable uncertainties in $\chi_0$ translate into modest uncertainties in $m_a$, also observed in~\cite{borsanyia}. Similarly the dependence on $n$ ($T_2$) is mild.  The lattice results of \cite{borsanyia,borsanyinature}, for example, differ from our estimate of $\chi_0(1.5)$ by a factor of about 10, and  exhibit power-law behavior corresponding to $n\simeq 8$.  If the Peccei-Quinn phase transition occurs after inflation, this factor of 10 leads to about a $15\%$ decrease in the value of the axion mass required to account
for the observed dark matter density.

To obtain a more accurate result for the late-time relic density, we solve the full axion equation of motion numerically through the time where it starts to oscillate. 
Fig.~\ref{fig:relic1} shows the relic density obtained in this way for two values of $T_2$ and a range of $\chi_0$, in the post-inflationary PQ-breaking scenario ($\theta_0=2.155$). Compared to the analytic estimate~(\ref{omegaresult}), the full numerical solution yields marginally higher $\Omega$ for fixed $m_a$. Even with the factor of $5$ variation in $T_2$ and the factor of $20^2$ variation in $\chi_0$, we find that the axion mass required to account for all of dark matter varies by only a factor of 2-3.

Additional sources of axion production (cosmic strings) 
can force a larger axion mass.  These masses are, indeed, at the edge of capability of cavity experiments like ADMX, and are the focus
of much future planning. However, these sources of energy density, as well as the constraint on $\theta_0$, are not necessarily present in the early universe.

\begin{figure}[t!]
\begin{center}
\includegraphics[width=0.45\linewidth]{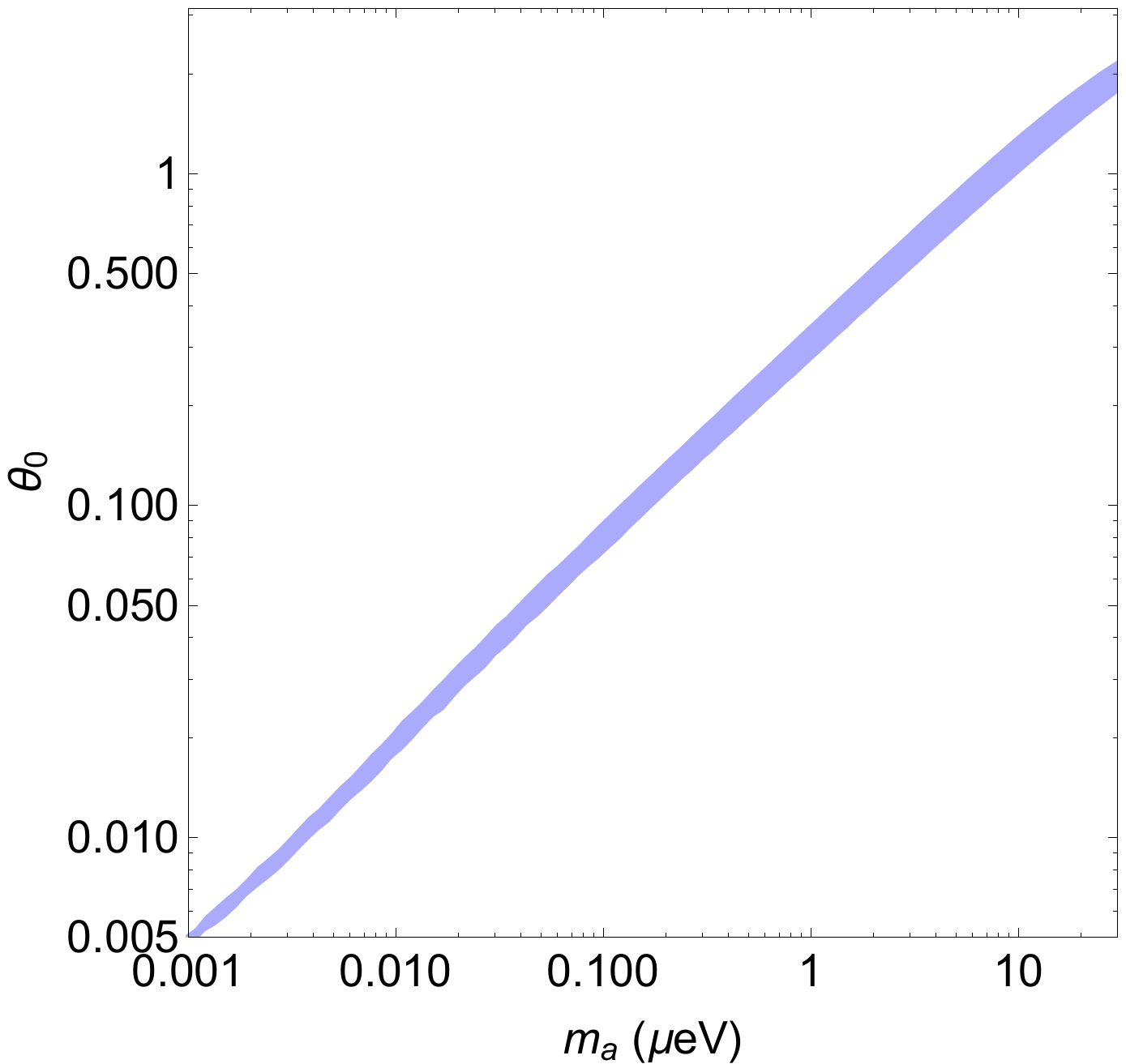}~~
\includegraphics[width=0.43\linewidth]{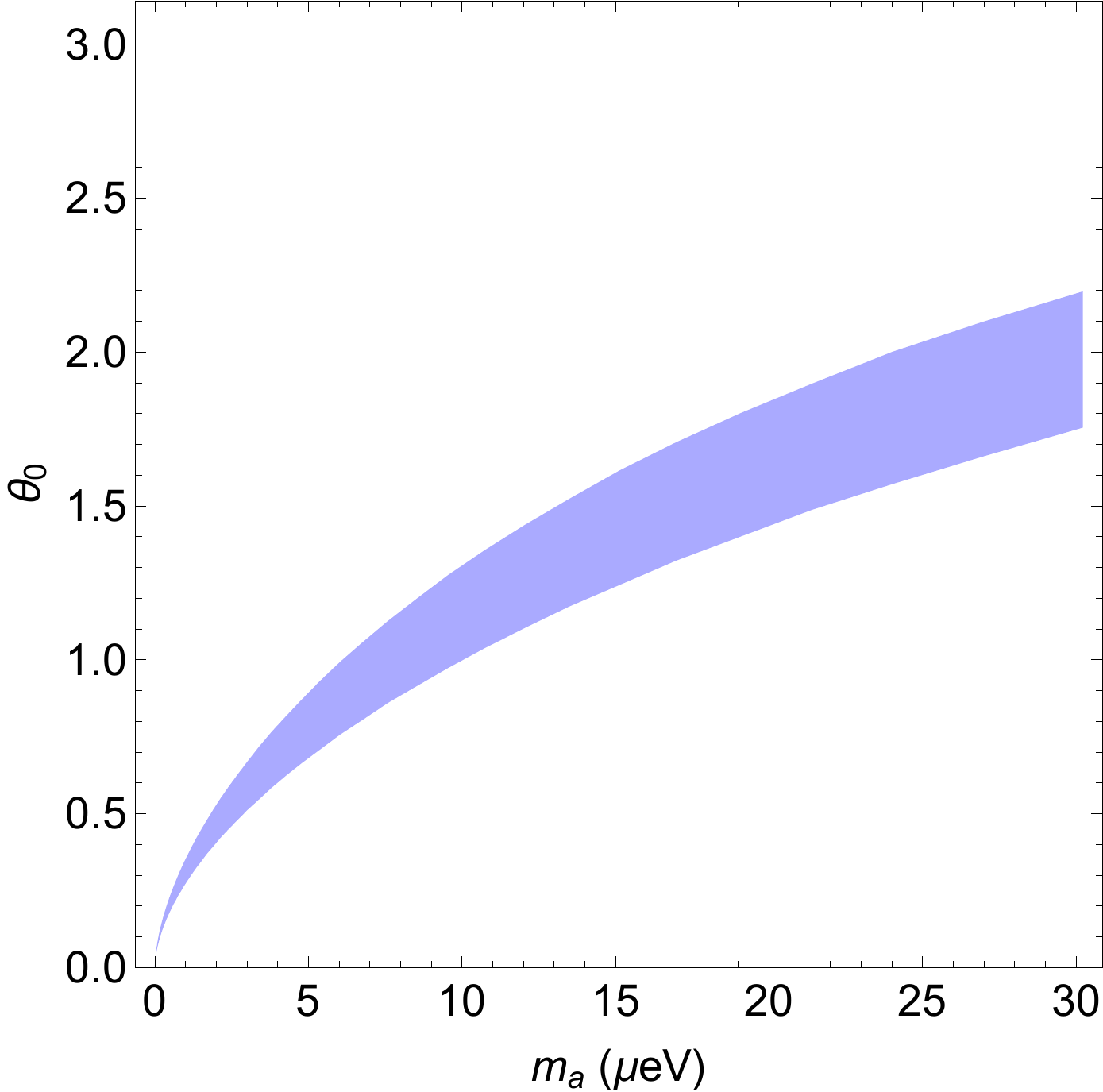}
\caption{Axion relic density from misalignment in the pre-inflationary scenario. The curve shows the misalignment angle needed to obtain $\Omega=0.258$.  The band reflects the uncertainty in the instanton computation of the free energy, Eq.~(\ref{eq:uncdeb}), used as an anchor at $T=1.5$ GeV, and the anchor point for ChPT has been fixed to $T_2=140$ MeV. Left panel: log-log axes over a broad range of axion masses. Right panel: linear axes over a range of axion masses in reach of current and next generation ADMX~\cite{admxfuture}.} 
\label{fig:preinflation}
\end{center}
\end{figure}

In fact, there is not necessarily a Peccei-Quinn transition at all~\cite{anisimovdine}.
The approximate Peccei-Quinn symmetry, if it exists, is almost certainly an accident.  This accident may not occur at the high temperatures or high curvatures that 
characterize the early universe.  In this case, the initial value of $\theta$ is a fixed number, or possibly one of a set of discrete numbers.  This number might well be small, or might be ${\cal{O}}(1)$.  Either has significant implications for the final dark matter density.

Alternatively, there may be an approximate symmetry both for low and high temperature (or curvature).  The question of whether the symmetry is broken during or after inflation then depends, for example, on the coupling of the inflaton to the
field responsible for PQ symmetry breaking.   For example, there might be an effective mass term for this field, of either sign.
There seems to be no particular reason to believe that one or the other outcome is favored.  

These different possibilities have been extensively studied in the literature. If the symmetry breaking occurs after inflation, one has to average over random initial misalignment angles, which fixes the parameter $\theta_0$ as above. 
In the case of symmetry breaking before inflation,
$\theta_0$ is a free parameter. In Fig.~\ref{fig:preinflation} we show the sensitivity of the misalignment angle required to saturate the relic density to the uncertainty in $\chi_0$. As in the post-inflationary case studied above, we find that the theoretical uncertainties have essentially no qualitative impact on the required parameters. Furthermore, for a wide range of ${\cal O}(1)$ values for $\theta_0$, the relevant axion masses are compatible with current and next-generation cavity experiments~\cite{Rosenberg:2015kxa,admxfuture}.

\section{Conclusions}
\label{conclusions}

Within the conventional picture of axion cosmology, we have found that the standard computation of the axion relic density is relatively robust against theoretical uncertainties stemming from the dilute gas computation of the QCD free energy at high temperatures and the behavior of the free energy at strong coupling. In particular, we have argued that the instanton computation is under sufficient control at temperatures of 1-2 GeV to allow a reasonable assessment of uncertainties due to higher-order corrections. These corrections cannot amount to much more than an order of magnitude in the free energy without an unexplained breakdown in the semiclassical analysis. In particular, we have argued that infrared divergences in the Debye mass in the perturbative vacuum are not relevant in the instanton background and cannot inject arbitrarily large corrections to $F(\theta,T)$. Thus, while an improved determination of the finite-temperature topological susceptibility would lead to improvement in the precision of the (relic density, axion mass) relation, it is not expected to lead to qualitative (order-of-magnitude) changes, and modern cavity experiments retain significant discovery potential. 

However, in closing, we note that it has long been recognized that the underlying cosmological assumptions of the standard calculation may not
hold, and that there is good theoretical motivation to consider lighter axions with larger decay constants.  Within conventional
effective field theory, for example, it is hard to account for the requisite quality of the Peccei-Quinn symmetry without invoking large discrete symmetries.  
String theory points to a different picture, in which the Peccei-Quinn symmetry
appears more natural~\cite{wittenaxion}.  The assumption that the underlying mass scale
is of order the Planck or unification scale is suggestive of larger decay constants.  
It could also be that four-dimensional effective
field theory is not useful at scales orders of magnitude below the Planck scale, as in large or warped extra dimension
scenarios, and early universe cosmology might be substantially modified.

\vskip 1cm
\noindent
%{\bf Acknowledgements:} 
\noindent
{\bf Acknowledgements:}  This work was supported in part by the U.S. Department of Energy grant number DE-FG02-04ER41286. 
We thank N. Arkani-Hamed and R. Kitano for valuable comments, and Leslie Rosenberg for bringing the work of~\cite{borsanyinature} to our
attention.  M.D. thanks the Institute for Advanced Study for its hospitality while much of this work was performed.

\bibliography{lattice_axions}{}

\providecommand{\href}[2]{#2}\begingroup\raggedright\begin{thebibliography}{10}

\bibitem{abbottsikivie}
L.~F. Abbott and P.~Sikivie, ``{A Cosmological Bound on the Invisible Axion},''
\href{http://dx.doi.org/10.1016/0370-2693(83)90638-X}{{\em Phys. Lett.}
  {\bfseries B120} (1983) 133--136}.
%%CITATION = PHLTA,B120,133;%%.

\bibitem{preskillwilczekwise}
J.~Preskill, M.~B. Wise, and F.~Wilczek, ``{Cosmology of the Invisible
  Axion},''
\href{http://dx.doi.org/10.1016/0370-2693(83)90637-8}{{\em Phys. Lett.}
  {\bfseries B120} (1983) 127--132}.
%%CITATION = PHLTA,B120,127;%%.

\bibitem{dinefischler}
M.~Dine and W.~Fischler, ``{The Not So Harmless Axion},''
\href{http://dx.doi.org/10.1016/0370-2693(83)90639-1}{{\em Phys. Lett.}
  {\bfseries B120} (1983) 137--141}.
%%CITATION = PHLTA,B120,137;%%.

\bibitem{Rosenberg:2015kxa}
L.~J. Rosenberg,
  \href{http://dx.doi.org/10.1073/pnas.1308788112}{``{Dark-matter QCD-axion
  searches},''} in {\em {Sackler Colloquium: Dark Matter Universe: On the
  Threshhold of Discovery Irvine, USA, October 18-20, 2012}}.
\newblock 2015.
\newblock
\url{http://arstechnica.com/science/2015/01/if-dark-matter-is-really-axions-we-could-find-out-soon/}.
\newblock
%%CITATION = INSPIRE-1355312;%%.

\bibitem{grahamrosenberg}
P.~W. Graham, I.~G. Irastorza, S.~K. Lamoreaux, A.~Lindner, and K.~A. van
  Bibber, ``{Experimental Searches for the Axion and Axion-Like Particles},''
  \href{http://dx.doi.org/10.1146/annurev-nucl-102014-022120}{{\em Ann. Rev.
  Nucl. Part. Sci.} {\bfseries 65} (2015) 485--514},
\href{http://arxiv.org/abs/1602.00039}{{\ttfamily arXiv:1602.00039 [hep-ex]}}.
%%CITATION = ARXIV:1602.00039;%%.

\bibitem{gpy}
D.~J. Gross, R.~D. Pisarski, and L.~G. Yaffe, ``{QCD and Instantons at Finite
  Temperature},''
\href{http://dx.doi.org/10.1103/RevModPhys.53.43}{{\em Rev. Mod. Phys.}
  {\bfseries 53} (1981) 43}.
%%CITATION = RMPHA,53,43;%%.

\bibitem{bonatia}
C.~Bonati, M.~D'Elia, M.~Mariti, G.~Martinelli, M.~Mesiti, F.~Negro,
  F.~Sanfilippo, and G.~Villadoro, ``{Axion phenomenology and
  $\theta$-dependence from $N_f = 2+1$ lattice QCD},''
  \href{http://dx.doi.org/10.1007/JHEP03(2016)155}{{\em JHEP} {\bfseries 03}
  (2016) 155},
\href{http://arxiv.org/abs/1512.06746}{{\ttfamily arXiv:1512.06746 [hep-lat]}}.
%%CITATION = ARXIV:1512.06746;%%.

\bibitem{borsanyia}
S.~Borsanyi, M.~Dierigl, Z.~Fodor, S.~D. Katz, S.~W. Mages, D.~Nogradi,
  J.~Redondo, A.~Ringwald, and K.~K. Szabo, ``{Axion cosmology, lattice QCD and
  the dilute instanton gas},''
  \href{http://dx.doi.org/10.1016/j.physletb.2015.11.020}{{\em Phys. Lett.}
  {\bfseries B752} (2016) 175--181},
\href{http://arxiv.org/abs/1508.06917}{{\ttfamily arXiv:1508.06917 [hep-lat]}}.
%%CITATION = ARXIV:1508.06917;%%.

\bibitem{berkowitz}
E.~Berkowitz, M.~I. Buchoff, and E.~Rinaldi, ``{Lattice QCD input for axion
  cosmology},'' \href{http://dx.doi.org/10.1103/PhysRevD.92.034507}{{\em Phys.
  Rev.} {\bfseries D92} no.~3, (2015) 034507},
\href{http://arxiv.org/abs/1505.07455}{{\ttfamily arXiv:1505.07455 [hep-ph]}}.
%%CITATION = ARXIV:1505.07455;%%.

\bibitem{kitanorapidfalloff}
R.~Kitano and N.~Yamada, ``{Topology in QCD and the axion abundance},''
  \href{http://dx.doi.org/10.1007/JHEP10(2015)136}{{\em JHEP} {\bfseries 10}
  (2015) 136},
\href{http://arxiv.org/abs/1506.00370}{{\ttfamily arXiv:1506.00370 [hep-ph]}}.
%%CITATION = ARXIV:1506.00370;%%.

\bibitem{kitano}
J.~Frison, R.~Kitano, H.~Matsufuru, S.~Mori, and N.~Yamada, ``{Topological
  susceptibility at high temperature on the lattice},''
  \href{http://dx.doi.org/10.1007/JHEP09(2016)021}{{\em JHEP} {\bfseries 09}
  (2016) 021},
\href{http://arxiv.org/abs/1606.07175}{{\ttfamily arXiv:1606.07175 [hep-lat]}}.
%%CITATION = ARXIV:1606.07175;%%.

\bibitem{sharma}
P.~Petreczky, H.-P. Schadler, and S.~Sharma, ``{The topological susceptibility
  in finite temperature QCD and axion cosmology},''
  \href{http://dx.doi.org/10.1016/j.physletb.2016.09.063}{{\em Phys. Lett.}
  {\bfseries B762} (2016) 498--505},
\href{http://arxiv.org/abs/1606.03145}{{\ttfamily arXiv:1606.03145 [hep-lat]}}.
%%CITATION = ARXIV:1606.03145;%%.

\bibitem{borsanyinature}
S.~Borsanyi {\em et~al.}, ``{Calculation of the axion mass based on
  high-temperature lattice quantum chromodynamics},''
  \href{http://dx.doi.org/10.1038/nature20115}{{\em Nature} {\bfseries 539}
  no.~7627, (2016) 69--71},
\href{http://arxiv.org/abs/1606.07494}{{\ttfamily arXiv:1606.07494 [hep-lat]}}.
%%CITATION = ARXIV:1606.07494;%%.

\bibitem{bonatib}
C.~Bonati, M.~D'Elia, M.~Mariti, G.~Martinelli, M.~Mesiti, F.~Negro,
  F.~Sanfilippo, and G.~Villadoro, ``{Recent progress on QCD inputs for axion
  phenomenology},'' \href{http://dx.doi.org/10.1051/epjconf/201713708004}{{\em
  EPJ Web Conf.} {\bfseries 137} (2017) 08004},
\href{http://arxiv.org/abs/1612.06269}{{\ttfamily arXiv:1612.06269 [hep-lat]}}.
%%CITATION = ARXIV:1612.06269;%%.

\bibitem{Trunin:2015yda}
A.~Trunin, F.~Burger, E.-M. Ilgenfritz, M.~P. Lombardo, and
  M.~Muller-Preussker, ``{Topological susceptibility from $N_f=2+1+1$ lattice
  QCD at nonzero temperature},''
  \href{http://dx.doi.org/10.1088/1742-6596/668/1/012123}{{\em J. Phys. Conf.
  Ser.} {\bfseries 668} no.~1, (2016) 012123},
\href{http://arxiv.org/abs/1510.02265}{{\ttfamily arXiv:1510.02265 [hep-lat]}}.
%%CITATION = ARXIV:1510.02265;%%.

\bibitem{Burger:2017xkz}
F.~Burger, E.-M. Ilgenfritz, M.~P. Lombardo, M.~Muller-Preussker, and
  A.~Trunin, ``{Topology (and axion's properties) from lattice QCD with a
  dynamical charm},'' in {\em {26th International Conference on
  Ultrarelativistic Nucleus-Nucleus Collisions (Quark Matter 2017)
  Chicago,Illinois, USA, February 6-11, 2017}}.
\newblock 2017.
\newblock \href{http://arxiv.org/abs/1705.01847}{{\ttfamily arXiv:1705.01847
  [hep-lat]}}.
\newblock
\url{http://inspirehep.net/record/1598136/files/arXiv:1705.01847.pdf}.
\newblock
%%CITATION = ARXIV:1705.01847;%%.

\bibitem{villadoro}
G.~Grilli~di Cortona, E.~Hardy, J.~Pardo~Vega, and G.~Villadoro, ``{The QCD
  axion, precisely},'' \href{http://dx.doi.org/10.1007/JHEP01(2016)034}{{\em
  JHEP} {\bfseries 01} (2016) 034},
\href{http://arxiv.org/abs/1511.02867}{{\ttfamily arXiv:1511.02867 [hep-ph]}}.
%%CITATION = ARXIV:1511.02867;%%.

\bibitem{rundec1}
K.~G. Chetyrkin, J.~H. Kuhn, and M.~Steinhauser, ``{RunDec: A Mathematica
  package for running and decoupling of the strong coupling and quark
  masses},'' \href{http://dx.doi.org/10.1016/S0010-4655(00)00155-7}{{\em
  Comput. Phys. Commun.} {\bfseries 133} (2000) 43--65},
\href{http://arxiv.org/abs/hep-ph/0004189}{{\ttfamily arXiv:hep-ph/0004189
  [hep-ph]}}.
%%CITATION = HEP-PH/0004189;%%.

\bibitem{rundec2}
B.~Schmidt and M.~Steinhauser, ``{CRunDec: a C++ package for running and
  decoupling of the strong coupling and quark masses},''
  \href{http://dx.doi.org/10.1016/j.cpc.2012.03.023}{{\em Comput. Phys.
  Commun.} {\bfseries 183} (2012) 1845--1848},
\href{http://arxiv.org/abs/1201.6149}{{\ttfamily arXiv:1201.6149 [hep-ph]}}.
%%CITATION = ARXIV:1201.6149;%%.

\bibitem{rundec3}
F.~Herren and M.~Steinhauser, ``{Version 3 of {\tt RunDec} and {\tt
  CRunDec}},''
\href{http://arxiv.org/abs/1703.03751}{{\ttfamily arXiv:1703.03751 [hep-ph]}}.
%%CITATION = ARXIV:1703.03751;%%.

\bibitem{morris}
T.~R. Morris, D.~A. Ross, and C.~T. Sachrajda, ``{Higher Order Quantum
  Corrections in the Presence of an Instanton Background Field},''
\href{http://dx.doi.org/10.1016/0550-3213(85)90131-2}{{\em Nucl. Phys.}
  {\bfseries B255} (1985) 115--148}.
%%CITATION = NUPHA,B255,115;%%.

\bibitem{rebhan}
A.~K. Rebhan, ``{The NonAbelian Debye mass at next-to-leading order},''
  \href{http://dx.doi.org/10.1103/PhysRevD.48.R3967}{{\em Phys. Rev.}
  {\bfseries D48} (1993) R3967--R3970},
\href{http://arxiv.org/abs/hep-ph/9308232}{{\ttfamily arXiv:hep-ph/9308232
  [hep-ph]}}.
%%CITATION = HEP-PH/9308232;%%.

\bibitem{arnoldyaffe}
P.~B. Arnold and L.~G. Yaffe, ``{The NonAbelian Debye screening length beyond
  leading order},'' \href{http://dx.doi.org/10.1103/PhysRevD.52.7208}{{\em
  Phys. Rev.} {\bfseries D52} (1995) 7208--7219},
\href{http://arxiv.org/abs/hep-ph/9508280}{{\ttfamily arXiv:hep-ph/9508280
  [hep-ph]}}.
%%CITATION = HEP-PH/9508280;%%.

\bibitem{creamer}
L.~S. Brown, R.~D. Carlitz, D.~B. Creamer, and C.-k. Lee, ``{Propagation
  Functions in Pseudoparticle Fields},''
\href{http://dx.doi.org/10.1103/PhysRevD.17.1583}{{\em Phys. Rev.} {\bfseries
  D17} (1978) 1583}.
%%CITATION = PHRVA,D17,1583;%%.

\bibitem{Levine:1978ge}
H.~Levine and L.~G. Yaffe, ``{HIGHER ORDER INSTANTON EFFECTS},''
\href{http://dx.doi.org/10.1103/PhysRevD.19.1225}{{\em Phys. Rev.} {\bfseries
  D19} (1979) 1225}.
%%CITATION = PHRVA,D19,1225;%%.

\bibitem{tcohen}
T.~D. Cohen, ``{The High temperature phase of QCD and U(1)-A symmetry},''
  \href{http://dx.doi.org/10.1103/PhysRevD.54.R1867}{{\em Phys. Rev.}
  {\bfseries D54} (1996) R1867--R1870},
\href{http://arxiv.org/abs/hep-ph/9601216}{{\ttfamily arXiv:hep-ph/9601216
  [hep-ph]}}.
%%CITATION = HEP-PH/9601216;%%.

\bibitem{aoki}
S.~Aoki, H.~Fukaya, and Y.~Taniguchi, ``{Chiral symmetry restoration,
  eigenvalue density of Dirac operator and axial U(1) anomaly at finite
  temperature},'' \href{http://dx.doi.org/10.1103/PhysRevD.86.114512}{{\em
  Phys. Rev.} {\bfseries D86} (2012) 114512},
\href{http://arxiv.org/abs/1209.2061}{{\ttfamily arXiv:1209.2061 [hep-lat]}}.
%%CITATION = ARXIV:1209.2061;%%.

\bibitem{Azcoiti:2016zbi}
V.~Azcoiti, ``{Topology in the SU(Nf) chiral symmetry restored phase of
  unquenched QCD and axion cosmology},''
  \href{http://dx.doi.org/10.1103/PhysRevD.94.094505}{{\em Phys. Rev.}
  {\bfseries D94} no.~9, (2016) 094505},
\href{http://arxiv.org/abs/1609.01230}{{\ttfamily arXiv:1609.01230 [hep-lat]}}.
%%CITATION = ARXIV:1609.01230;%%.

\bibitem{Azcoiti:2017jsh}
V.~Azcoiti, ``{Topology in the SU($N_f$) chiral symmetry restored phase of
  unquenched QCD and axion cosmology II},''
\href{http://arxiv.org/abs/1704.04906}{{\ttfamily arXiv:1704.04906 [hep-lat]}}.
%%CITATION = ARXIV:1704.04906;%%.

\bibitem{turnerreview}
M.~S. Turner, ``{Windows on the Axion},''
\href{http://dx.doi.org/10.1016/0370-1573(90)90172-X}{{\em Phys. Rept.}
  {\bfseries 197} (1990) 67--97}.
%%CITATION = PRPLC,197,67;%%.

\bibitem{kolbturner}
E.~W. Kolb and M.~S. Turner, ``{The Early Universe},''
{\em Front. Phys.} {\bfseries 69} (1990) 1--547.
%%CITATION = FRPHA,69,1;%%.

\bibitem{admxfuture}
\url{http://depts.washington.edu/admx/future.shtml}.

\bibitem{anisimovdine}
M.~Dine and A.~Anisimov, ``{Is there a Peccei-Quinn phase transition?},''
  \href{http://dx.doi.org/10.1088/1475-7516/2005/07/009}{{\em JCAP} {\bfseries
  0507} (2005) 009},
\href{http://arxiv.org/abs/hep-ph/0405256}{{\ttfamily arXiv:hep-ph/0405256
  [hep-ph]}}.
%%CITATION = HEP-PH/0405256;%%.

\bibitem{wittenaxion}
E.~Witten, ``{Some Properties of O(32) Superstrings},''
\href{http://dx.doi.org/10.1016/0370-2693(84)90422-2}{{\em Phys.Lett.}
  {\bfseries B149} (1984) 351--356}.
%%CITATION = PHLTA,B149,351;%%.

\end{thebibliography}\endgroup
\bibliographystyle{utphys}
%\bibliographystyle{unsrt}
%\bibliographystyle{JHEP}
%\bibliography{Biblio}

\end{document}